\documentclass[aps,twocolumn,floats,amsmath,amssymb]{revtex4}

\usepackage{graphics}
\usepackage{dcolumn}

\begin{document}

\title{{\rm\small\hfill (submitted to J. Phys. Condens. Matter)}\\
O- and H- induced surface core level shifts on Ru(0001):\\
Prevalence of the additivity rule}

\author{S. Lizzit$^1$, Y. Zhang$^2$, K.L. Kostov$^3$, L. Petaccia$^1$, A. Baraldi$^{4,5}$,\\
D. Menzel$^{2,6}$ and K. Reuter$^2$}

\address{$^1$Sincrotrone Trieste, Strada Statale 14 Km 163.5, 34012 Trieste, Italy}

\address{$^2$Fritz-Haber-Institut der MPG, Faradayweg 4-6, 14195 Berlin, Germany}

\address{$^3$Institute of General and Inorganic Chemistry, Bulgarian Academy of Science, 1113 Sofia, Bulgaria}

\address{$^4$ Physics Department and Center of Excellence for Nanostructured Materials, University of Trieste, via A Valerio 2, 34127 Trieste, Italy}

\address{$^5$ Laboratorio Nazionale TASC INFM-CNR, in AREA Science Park, 34012 Trieste, Italy}

\address{$^6$Physik-Department E20, Technische Universit\"at M\"unchen, 85748 Garching, Germany}

\received{14th October 2008}

\begin{abstract}
In previous work on adsorbate-induced surface core level shifts (SCLSs), the effects caused by O atom adsorption on Rh(111) and Ru(0001) were found to be additive: the measured shifts for first layer Ru atoms depended linearly on the number of directly coordinated O atoms. Density-functional theory calculations quantitatively reproduced this effect, allowed separation of initial and final state contributions, and provided an explanation in terms of a roughly constant charge transfer per O atom. We have now conducted similar measurements and calculations for three well-defined adsorbate and coadsorbate layers containing O and H atoms: $(1 \times 1)$-H, $(2 \times 2)$-(O+H), and $(2 \times 2)$-(O+3H) on Ru(0001). As H is stabilized in fcc sites in the prior two structures and in hcp sites in the latter, this enables us to not only study coverage and coadsorption effects on the adsorbate-induced SCLSs, but also the sensitivity to similar adsorption sites. Remarkably good agreement is obtained between experiment and calculations for the energies and geometries of the layers, as well as for all aspects of the SCLS values. The additivity of the next-neighbor adsorbate-induced SCLSs is found to prevail even for the coadsorbate structures. While this confirms the suggested use of SCLSs as fingerprints of the adsorbate configuration, their sensitivity is further demonstrated by the slightly different shifts unambiguously determined for H adsorption in either fcc or hcp hollow sites.
\end{abstract}

\maketitle

\section{Introduction}

Core electron binding energies of solids are sensitive probes of the local atomic and geometric environment and as such markedly changed at surfaces. A quantitative understanding of the corresponding shifts, in particular of surface atoms coordinated to atomic or molecular adsorbates, provides a wealth of information e.g. on the nature of the chemical bond \cite{martensson94} and may even be used for the interpretation and prediction of surface reactions \cite{baraldi04}. Experimentally, the shifts can be accessed by high-resolution X-ray photoelectron spectroscopy (XPS), as part of the so-called surface core level shifts (SCLSs) \cite{martensson94}. Apart from the actual so-called initial-state shifts due to the changed electron density viz. electrostatic potential at the location of the core electrons, the measured SCLSs also contain so-called final-state contributions from variations in the screening of the core hole created in the XPS process, which is influenced by the surface or nearby adsorbates as well. At present, a partitioning of the experimentally only accessible total shifts into initial- and final-state effects is only possible by first-principles electronic structure calculations. In particular for metal surfaces, approaches based on density-functional theory (DFT) have recently proven to be very successful. For instance, remarkably quantitative agreement could be achieved with the SCLS measurements of the first-layer atoms of two close-packed transition metal surfaces, Rh(111) \cite{pirovano01} and Ru(0001) \cite{lizzit01}, and of their modification by varying amounts of adsorbed O atoms in well-defined structures. Moreover, these calculations were then able to differentiate charge density and screening contributions and to analyse the respective physical basis in terms of the structures and the electronic properties of the systems studied.

Especially for adsorbate systems the high-resolution XPS spectra generally exhibit multi-peaked structures arising from different environments of the surface substrate atoms (first, second, or deeper layers; neighbors to adsorbates of different type and number). Some assignment can often be reached by suitable spectroscopic means or by systematic variation and comparison of various sample conditions e.g. in form of different adsorbate coverages. However, a comprehensive interpretation requires detailed DFT calculations which in turn necessarily need to be based on a microscopic structural model. In view of a potential application to less well characterized systems more general relations enabling a direct data interpretation are therefore of interest. In this respect, the mentioned SCLS work at Rh(111) \cite{pirovano01} and Ru(0001) \cite{lizzit01} identified an intriguing scaling in that the O-induced SCLSs of the first-layer metal atoms turned out to be linearly proportional to the number of directly coordinated O atoms to good accuracy. With the induced SCLS containing only a small and not much coverage dependent final-state contribution, the source of this additivity could be traced back to a roughly constant amount of charge that is withdrawn from the surface atoms by each directly coordinated O neighbor \cite{lizzit01}. With ensuing work on Rh(100) confirming this then termed "additivity rule", it was suggested that the latter may be a universal property of SCLSs which could be used for the fingerprinting of adsorbate populations in catalytic applications \cite{baraldi04}.

The cases mentioned all concern O atoms, a species with high electronegativity. SCLS results have also been obtained for H atoms on Rh(111), (100), and (110), including coverage dependencies \cite{baraldi08}, but no results exist for mixed layers. It is therefore very interesting to examine whether such an additivity rule applies also to combinations of different adatoms. To this end we have selected H atom adsorption and O + H coadsorption on Ru(0001). This choice has the advantage that not only O \cite{lizzit01,lindroos89} and H \cite{lindroos87} atomic overlayers, but also two well-defined (O+H) coadsorbate systems \cite{schiffer00,gsell08} have already been characterized in detail in terms of adsorbate phases and geometries. On this basis we have measured the SCLS spectra for these systems and combined them with DFT calculations for ground state energies, geometries, electronic structure, and SCLS values for the various distinguishable Ru surface atoms. We find excellent agreement between experiment and theory both for the geometries of the stable structures and for the SCLSs of all surface atoms. The central result we obtain is that we find the additivity rule to apply quite well also for H adsorbate atoms, and even for coadsorbate structures containing the strongly charge-transfer inducing O atom and the roughly neutral H atom.

The paper is structured in the following way. An experimental section briefly summarizes the technique, the studied layers and their preparation, and displays the accumulated XPS data and the derived shift values. The fitting procedures to extract the latter from the measured spectra are briefly described in an Appendix. The next section details the computational procedures and compares the obtained stable geometries in detail with existing experimental data. We then compare the total calculated shifts to the experimental ones, and decompose these total SCLSs into initial- and final-state contributions. Concluding with a thorough discussion and analysis of the physical properties behind the observed ``additivity rule'', we can trace the latter again back to an almost constant $-$ albeit compared to O neighbors much smaller $-$ amount of charge transferred to each directly coordinated H neighbor.

\section{Experiment}
\subsection{Structures: Geometry and preparation}

\begin{figure}
\begin{center}
\scalebox{0.3}{\includegraphics{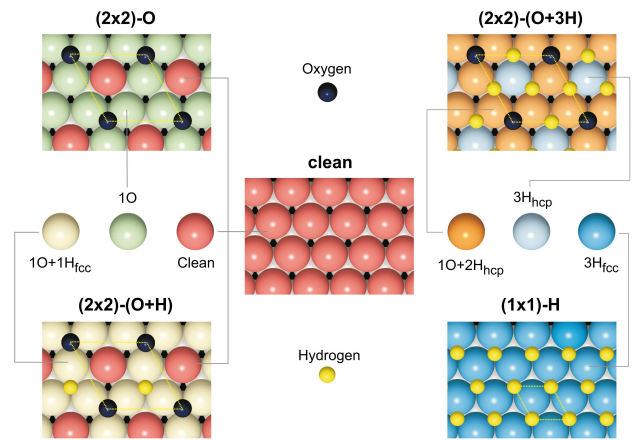}}
\end{center}
\caption{\label{fig1}
Schematic top views of the examined surface structures prepared starting from either the clean or $(2 \times 2)$-O covered Ru(0001) surface. The color-coding shows the different next-neighbor surroundings of the surface Ru atoms.}
\end{figure}

The experimental procedures were similar to those of ref. \cite{lizzit01}. The SuperESCA beamline of the synchrotron radiation source ELETTRA in Trieste was used which works at base pressures in the low 10$^{-10}$ mbar region and contains preparation facilities and an electron spectrometer with multichannel detection. The Ru(0001) surface was prepared by established procedures \cite{lizzit01,lindroos89,lindroos87,gsell08}; cleanliness and order of the surface were checked by XPS and low-energy electron diffraction (LEED). The following adsorbate structures have been investigated: $(1 \times 1)$-H, $(2 \times 2)$-O, $(2 \times 2)$-(O+H), and $(2 \times 2)$-(O+3H). For all structures, the adsorption sites and detailed geometries are known from previous quantitative IV-LEED or VLEED analysis \cite{lindroos89,gsell08}. Figure~\ref{fig1} compiles the structures and the sequence of preparation. In $(1 \times 1)$-H all hydrogen atoms sit in fcc sites. To prepare this saturated layer, the well-cleaned surface has to be exposed to high doses (several 1000 Langmuir) of molecular hydrogen, because the last few percent to saturation coverage adsorb very slowly \cite{kostov04}. Low background pressure when hydrogen is introduced is important; otherwise the layer becomes contaminated with water \cite{kostov04}. For the two O+H coadsorbate structures, the starting point is a well ordered $(2 \times 2)$-O structure with O adsorbed in hcp sites. The latter can be produced by slight overdosing with oxygen at 300 K to produce an O coverage of about 0.3 monolayer, heating to 1250 K, then heating to 400 K in 10$^{-7}$ mbar hydrogen for about 10 minutes, and finally heating to 400 K for ordering \cite{koch96}. Saturating this structure with hydrogen at 90 K, followed by short anneal at 100-110 K for ordering, produces the $(2 \times 2)$-(O+3H) structure. Upon heating the latter to 155 K with 2 K/s, or faster to 160 K, two thirds of the hydrogen is desorbed and the $(2 \times 2)$-(O+H) structure remains, as was proven by thermal desorption spectra (TPD) \cite{schiffer00} and quantitative LEED \cite{gsell08}. Quantitative LEED has also corroborated expectations from quantitative TPD \cite{schiffer00} that in the (O+3H) structure all H atoms sit in hcp sites, while in the (O+H) all H atoms sit in fcc sites \cite{gsell08}. This site selectivity in the two structures has been ascribed to repulsive O-H lateral interactions coupled with the tendency to maximize the coverage and we come back to this point below.

The interesting gist for the SCLS measurements is that the various structures contain Ru first layer atoms in a variety of surroundings as also schematized in Fig.~\ref{fig1}: with 3~${\rm H}_{\rm fcc}$ neighbors in the $(1 \times 1)$-H layer; with 1~O or without adsorbate neighbor in the $(2 \times 2)$-O; with 1~O + 2~${\rm H}_{\rm hcp}$ or 3~${\rm H}_{\rm hcp}$ in the $(2 \times 2)$-(O+3H); and with 1~O + 1~${\rm H}_{\rm fcc}$ or without adsorbate neighbor in the $(2 \times 2)$-(O+H). Together with the structures containing zero- to threefold O-coordinated Ru surface atoms studied in ref.~\cite{lizzit01}, this variety offers a good basis to systematically check on the SCLSs additivity with and without coadsorbates, as well as with H adsorbates in slightly different hollow sites.

\subsection{SCLS measurements and analysis}

\begin{figure}
\begin{center}
\scalebox{0.4}{\includegraphics{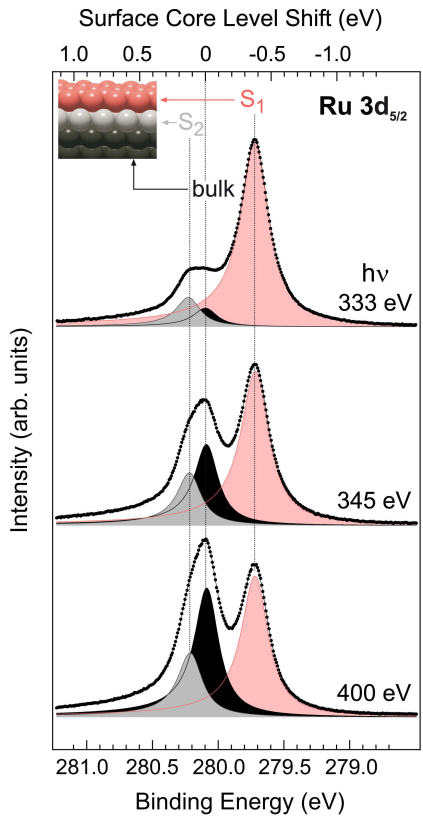}}
\end{center}
\caption{\label{fig2}
High-resolution XPS spectra of the Ru $3d_{5/2}$ core level of clean Ru(0001), measured (from top) at 333, 345, and 400 eV photon energy. The main components are marked (S$_1$: first layer Ru atoms; S$_2$: second layer Ru atoms; bulk Ru atoms). The bulk value is used as zero reference for the surface core level shifts (top $x$-axis).}
\end{figure}

The SCLS experiments were performed at three different photon energies (333, 345, and 400 eV) at a photon beam incidence angle of 70$^\circ$, which in the present setup leads to electron emission into the surface normal. Through spectroscopic means and detailed DFT calculations ref. \cite{lizzit01} has already established the assignment of the peaks to the different Ru atoms (first layer, second layer, bulk; modification by O neighbors) for the clean and the $(2 \times 2)$-O surface, thereby also facilitating the detection and analysis of new peaks in the H (co)adsorbate structures. As illustrated by Fig.~\ref{fig2} for clean Ru(0001) the data obtained at 333 eV photon energy appears most surface-sensitive in that both the first layer (S$_1$) and second layer (S$_2$) surface peaks dominate over the bulk intensity. All SCLS data and spectra shown in the following therefore correspond to this photon energy. The data acquired for the other energies has been analyzed as well, without obtaining significantly different findings to those detailed below. 

\begin{figure}
\begin{center}
\scalebox{0.4}{\includegraphics{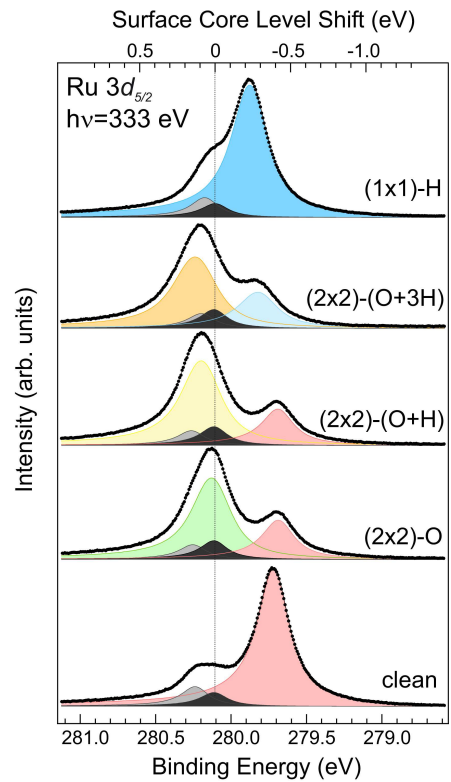}}
\end{center}
\caption{\label{fig3}
Ru $3d_{5/2}$ SCLS spectra for all studied structures. Additionally shown is the decomposition into the different peaks, where the color-coding indicates the assignment to the different surface atoms as explained in Fig. \ref{fig1}.}
\end{figure}

Figure~\ref{fig3} collects the SCLS measurements of all surfaces studied and shows the components derived from the fits; the numerical values are given in Table~3 ({\em vide infra}). The fitting procedures used to distinguish and disentangle the different components in the spectra are identical to those discussed in ref. \cite{lizzit01}, with further details given in the Appendix. The best-fit parameters (asymmetry $\alpha$, Lorentzian width $L$ and Gaussian width $G$) determined for the clean and $(2 \times 2)$-O surface are close to those published in ref. \cite{lizzit01}; small changes are explainable by the higher surface intensities obtained in the present measurements. In fact, we believe that the present values for the surface peaks are the more reliable ones. In any case, the central point is that all re-measured SCLSs for the clean and the $(2 \times 2)$-O surface are within error bars the same as in the previous work. The best-fit parameters determined for the H-containing structures are again similar, with further details given in the Appendix. One notable difference is that the components corresponding to Ru atoms with H neighbors had to be given somewhat larger widths, which we interpret by inhomogeneous broadening due to a less than perfect ordering achieved for these structures.

\section{Theory}
\subsection{Computational setup and SCLS calculations}

The DFT calculations were performed using the full-potential augmented plane wave plus local orbital (L)APW+lo method \cite{sjostedt00,madsen01} as implemented in the WIEN2k code \cite{wien2k}. The exchange-correlation (xc) energy is treated with the generalized gradient approximation (GGA) using the PBE xc-functional \cite{perdew96}. The procedure to obtain initial- and final-state contributions to the SCLSs is exactly the same as detailed in ref. \cite{lizzit01}, in which the initial-state shifts are simply given by Kohn-Sham eigenvalue differences and the final-state screening is computed within the Slater-Janak transition state approach by an impurity calculation with half an electron removed from the core state considered.

All surfaces are modeled with supercell geometries, using inversion symmetric slabs consisting of eight Ru(0001) layers with O or H adsorption on both sides. The vacuum between consecutive slabs is at least 11\,{\AA}. Within the employed $(2 \times 2)$ surface unit-cells the positions of all adsorbates and Ru atoms in the topmost three surface layers were fully relaxed for all structures considered. The gas-phase calculations for atomic H and O required for the binding energies are done in rectangular supercells with side lengths $(13 \times 14 \times 15)$\,bohr. The obtained total energies are then corrected by the gas-phase O$_2$ \cite{kiejna06} and H$_2$ binding energies to obtain adsorbate binding energies with respect to the molecular references. The basis set parameters are briefly summarized as follows: The muffin-tin spheres for Ru, O and H are $R_{\rm MT}^{\rm Ru} = 2.0$\,bohr, $R_{\rm MT}^{\rm O} = 1.0$\,bohr, and $R_{\rm MT}^{\rm H} = 0.8$\,bohr, respectively. Inside the muffin-tins the wave functions are expanded up to $l^{\rm wf}_{\rm max} = 12$ and the potential up to $l^{\rm pot}_{\rm max} = 6$. The energy cutoff for the expansion of the wave function in the interstitial is $E^{\rm wf}_{\rm max} = 20$\,Ry and for the potential $E^{\rm pot}_{\rm max} = 196$\,Ry. For the $(2 \times 2)$ surface unit-cells, a $(6 \times 6 \times 1)$ Monkhorst-Pack grid was used for the Brillouin zone integrations, for the gas-phase calculations in the rectangular supercells $\Gamma$-point sampling was employed.
 
The chosen computational approach is thus highly similar, if not superior to the one used already in ref. \cite{lizzit01}, with the notable exception that the now computationally feasible thicker slabs should enable a better description of the SCLSs of second layer Ru atoms. Recomputing all SCLSs of all oxygen-containing structures reported in ref. \cite{lizzit01} with the new computational setup, it is thus not surprising that we obtain with one exception values that are identical to the published ones to within 30 meV, thereby simply reconfirming the stringent convergence tests performed for the previous work. As expected the sole disagreement concerns a second layer shift in the $(2 \times 1)$-O structure, denoted as S$_2$(O) in ref. \cite{lizzit01}, that was previously computed as $-21$\,meV and is now obtained as $+61$\,meV. This does not affect the published comparison to the experimental data though, as the latter peak was not resolved in the measurements.

\subsection{Stable adsorption sites and adsorption geometries}

\begin{table*}
\caption{\label{table1}
Calculated total binding energy in meV and using gas-phase O$_2$ and H$_2$ as zero reference. Compared are the structural models for the $(2 \times 2)$-(O+H) and $(2 \times 2)$-(O+3H) coadsorption phases considered in ref. \cite{gsell08}, using exactly the numbering employed in this preceding work. Model-X means that the structure relaxed to the structure of model-X. The most stable structure corresponding to the highest binding energy (marked in bold) is for both phases exactly the one identified in the quantitative LEED analysis of ref. \cite{gsell08}, and is shown in Fig. \ref{fig1}.}
\begin{tabular}{@{}lcccccccc@{}}
\hline
                            & {\bf 1}    & 2    & 3    & 4       & 5        & 6        & 7       & 8    \\ \hline
$E_{\rm b,tot}^{\rm (O+H)}$ & {\bf 3400} & 3146 & 3301 & model-2 & model-3  & model-2  & model-1 & 3270 \\  
$E_{\rm b,tot}^{\rm (O+3H)}$& {\bf 3825} & 3049 & 3743 & 3752    & 2097     & model-1  & model-1 & model-1     \\ 
\end{tabular}
\end{table*}

In order to complement the preceding structure determination by quantitative LEED \cite{lindroos89,gsell08} a first series of calculations addressed the stable O and H adsorption sites by comparing the binding energy at the high-symmetry sites offered by the Ru(0001) surface. Consistent with previous calculations \cite{reuter02} and in full agreement to the experimental analysis, the most favorable O adsorption site in purely oxygen-containing $(2 \times 2)$ overlayers with one up to four O atoms per surface unit-cell is the hcp hollow site. On the contrary, for H atoms in the same $(2 \times 2)$ overlayers the most stable site is the fcc hollow site, albeit with only a small and largely coverage-independent difference of about 50\,meV/H atom to the hcp site. Concerning the $(2 \times 2)$-(O+H) and $(2 \times 2)$-(O+3H) coadsorption phases, we subjected all structural models considered in the recent quantitative LEED analysis of ref. \cite{gsell08} to a full geometry optimization. The results are summarized in Table~\ref{table1} and are consistent with the preceding work in that the model with the lowest Pendry $R_p$-factor is also the one with the highest binding energy and the detailed geometric parameters of these two phases summarized in Table~\ref{table2} agree very well. This nicely confirms the assessment of ref. \cite{gsell08} that LEED is indeed sensitive to the position of the H adsorbates. In these most stable structures the H atoms sit therefore, as already indicated in Fig.~\ref{fig1}, in the fcc sites in the $(2 \times 2)$-(O+H) configuration, and in the hcp sites in the $(2 \times 2)$-(O+3H) configuration.

\begin{table}
\caption{\label{table2}
Computed geometric parameters of the stable $(2 \times 2)$-(O+H) and $(2 \times 2)$-(O+3H) structures shown in Fig. \ref{fig1}, compared to the results of the experimental LEED structure determination of ref. \cite{gsell08}. $d$ is the adatom-Ru bond length, $d_{ij}$ the mean vertical distance between layer $i$ and $j$ (with the adsorbate forming layer 0 and the topmost Ru layer forming layer 1 etc.), and $z_i$ is the vertical buckling of layer $i$. For the mean vertical layer distances, the values in brackets additionally indicate the relative changes with respect to the bulk interlayer distance.}
\begin{tabular}{@{}lcc@{}}
\hline
& \multicolumn{2}{c}{$(2 \times 2)$-(O+H)} \\
	                          & Theory          & Experiment                     \\ \hline
$d$(H-Ru) (\AA)             & 1.89            & 1.83$\pm$0.13                  \\
$d$(O-Ru) (\AA)             & 2.01		        & 		                           \\
$\Delta d_{01}$(H-Ru) (\AA) & 1.15            & 1.03+0.15/-0.08                \\
$\Delta d_{01}$(O-Ru) (\AA) & 1.22            & 1.19$\pm$0.02                  \\
$\Delta d_{12}$ (\AA)       & 2.13 ($-$1.2\%) & 2.10$\pm$0.02 ($-1.9\pm0.9$\%) \\
$\Delta d_{23}$ (\AA)       & 2.18 ($+$1.5\%) & 2.14$\pm$0.02 ($0.0\pm0.9$\%)  \\
$\Delta d_{34}$ (\AA)       & 2.17 ($+$1.0\%) &                                \\
$\Delta z_{1}$ (\AA)        & 0.1             & 0.1                            \\
$\Delta z_{2}$ (\AA)        & 0.05            & 0.01                           \\
$\Delta z_{3}$ (\AA)        & 0.03            &                                \\[0.5cm] \hline

& \multicolumn{2}{c}{$(2 \times 2)$-(O+3H)} \\
                          	& Theory          & Experiment                     \\ \hline
$d$(H-Ru) (\AA)             & 1.86 $-$ 1.88   & 1.89$\pm$0.3                   \\
$d$(O-Ru) (\AA)             & 2.0			        &                                \\
$\Delta d_{01}$(H-Ru) (\AA) & 1.02		        & 1.08$\pm$0.15                  \\
$\Delta d_{01}$(O-Ru) (\AA) & 1.24	          & 1.19$\pm$0.025                 \\
$\Delta d_{12}$ (\AA)       & 2.18 (+1.2\%)   & 2.14$\pm$0.02 ($0.0\pm0.9$\%)  \\
$\Delta d_{23}$ (\AA)       & 2.16 (+0.3\%)   & 2.12$\pm$0.02 ($-0.9\pm0.9$\%) \\
$\Delta d_{34}$ (\AA)       & 2.18 (+1.2\%)   &                                \\
$\Delta z_{1}$ (\AA)        & 0.09            &  0.07                          \\
$\Delta z_{2}$ (\AA)        & 0.02            &  0.03                          \\
$\Delta z_{3}$ (\AA)        & 0.03            &                                \\
\end{tabular}
\end{table}

\section{SCLS results and discussion}
\subsection{Comparison of computed and measured SCLSs}

From the geometric structures shown in Fig.~\ref{fig1} one expects the following first layer components in the measured SCLS spectra: a peak due to Ru atoms without adsorbate coordination (S$_1$) in the clean surface, as well as peaks due to zero- and onefold O-coordinated Ru atoms in the $(2 \times 2)$-O structure, S$_1$ and S$_1$(O) respectively. In the $(2 \times 2)$-(O+H) structure there are Ru atoms coordinated to 1~O + 1~${\rm H}_{\rm fcc}$ (S$_1$(O+H$_{\rm fcc}$)) and without adsorbate neighbor (S$_1$); in the $(2 \times 2)$-(O+3H) structure there are Ru atoms coordinated to 1~O + 2~${\rm H}_{\rm hcp}$ (S$_1$(O+2H$_{\rm hcp}$)) and Ru atoms coordinated to 3~${\rm H}_{\rm hcp}$ (S$_1$(3H$_{\rm hcp}$); and in the $(1 \times 1)$-H layer there are Ru atoms coordinated to 3~${\rm H}_{\rm fcc}$ neighbors (S$_1$(3H$_{\rm fcc}$)). In addition there can be non-negligible second layer shifts (S$_2$), where in the $(2 \times 2)$ layers one needs to furthermore distinguish between Ru atoms that are directly underneath an adsorbed O atom (S$_2$(O)) and those that are not (S$_2$). The SCLS data analysis and assignment for the clean surface and $(2 \times 2)$-O layer has already been established in ref. \cite{lizzit01} and is the basis for the decomposition of the now measured spectra shown in Fig.~\ref{fig3}. Specifically, we followed the strategy explained in detail in ref. \cite{lizzit01} in which no attempt is made to resolve the small S$_2$(O) second layer component in the O-containing structure. Within this approach the assignment of the new peaks in the H overlayer and the two H-containing coadsorbate structures is straightforward. Also for the latter two only one non-negligible second layer peak was considered and we come back to the consequences for the derived SCLS values below. However, we note already here that although small, the shift of the first layer S$_1$(O) component of the $(2 \times 2)$-O overlayer when it becomes the S$_1$(O+H$_{\rm fcc}$) component in the $(2 \times 2)$-(O+H) phase and even further when it is turned into the S$_1$(O+2H$_{\rm hcp}$) component in the $(2 \times 2)$-(O+3H) phase is directly visible from the spectra and thus independent of the specific decomposition procedure employed. The same holds for the shift of the S$_1$ peak of the clean surface when turned into the S$_1$(3H$_{\rm fcc}$) component of the $(1 \times 1)$-H layer, cf. Fig.~\ref{fig3}.

\begin{table*}
\caption{\label{table3}
Comparison of measured and calculated SCLS values for the five investigated structures. All shifts are given in meV. See text for an explanation of the employed nomenclature. No attempt was made to fit a S$_2$(O) component in the experimental spectra (see text).}
\begin{tabular}{@{}l@{}|@{}ccccc@{}}
\hline
Exp./Theory   & Clean      & $(2\!\times\!2)$-O & $(2\!\times\!2)$-(O+H) & $(2\!\times\!2)$-(O+3H) & $(1\!\times\!1)$-H \\ \hline
S$_1$            & $-$371/$-$405 & $-$413/$-$423    & $-$414/$-$438        &                         & \\
S$_1$(3H$_{\rm hcp})$  &      &                  &                      & $-$291/$-$331           & \\
S$_1$(3H$_{\rm fcc})$  &      &                  &                      &                         & $-$210/$-$205 \\
S$_1$(O)         &            & +24/+29          &                      &                         & \\
S$_1$(O+H$_{\rm fcc}$)&       &                  & +89/+91              &                         & \\
S$_1$(O+2H$_{\rm hcp}$)&      &                  &                      & +120/+120               & \\ \hline
S$_2$            & +133/+102  & +144/+155        & +135/+158            & +85/+54                 & +78/+52 \\
S$_2$(O)         &            & n.a./$-$32       & n.a./$-$20           & n.a./$-$22              & \\
\end{tabular}
\end{table*}

\begin{figure}
\begin{center}
\scalebox{0.5}{\includegraphics{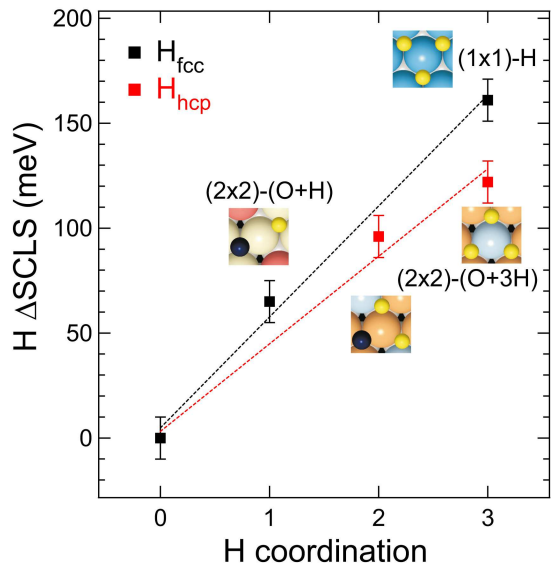}}
\end{center}
\caption{\label{fig4}
Experimental H-induced changes of the first layer SCLSs as a function of the number of directly coordinated H atoms. Shown is the value by which the SCLS changes upon H neighbor addition, i.e. the reference is the S$_1$ value of the clean Ru(0001) surface for the purely H-coordinated Ru atoms, the S$_1$ value of the $(2 \times 2)$-O structure for the Ru atoms coordinated solely to H in the coadsorbate phases, and the S$_1$(O) value of the $(2 \times 2)$-O structure for the Ru atoms coordinated to both O and H in the coadsorbate phases.}
\end{figure}

Table~\ref{table3} compiles the SCLS values derived from the decomposition shown in Fig.~\ref{fig3} and compares them to the calculated total shifts. The agreement is on the same remarkable level as that of ref. \cite{lizzit01}, with maximum deviations of the order of 40\,meV. As discussed in detail in this preceding work, one could presumably reach an even better agreement by explicitly considering the small S$_2$(O) component in the experimental spectra, which is, however, not necessary for the scope of this work. Essentially, the experiment-theory comparison is thus flawless, which reconfirms not only that the employed GGA-PBE functional seems to allow for a quantitative determination of the quantity of interest to our study, but also fully supports the experimental peak assignment. On this basis it is intriguing to realize that the derived first layer SCLS values are again quasi-linearly additive with respect to the coordination with H atoms. This is better visualized in Fig. \ref{fig4} which shows the H-induced changes of the SCLSs, i.e. the amount by which a first layer Ru SCLS changes upon adding another H nearest neighbor. Strikingly, the observed additivity seems even independent of the presence of the O coadsorbate, and slightly different shifts are obtained for H neighbors in either fcc or hcp hollow sites. We furthermore mention that the O- and H-induced shifts are also closely the same to the ones observed in the $(2 \times 2)$-(O+H) structure for a $(2 \times 2)$-(O+H+CO) structure which additionally contains CO on top of those Ru surface atoms, which in the $(2 \times 2)$-(O+H) have no direct adsorbate neighbors \cite{sclstobepublished}. This suggests also a minor influence of near-next neighbor occupation compared to the here discussed direct neighbor effects.

\subsection{Analysis of the ``additivity rule''}

\begin{table}
\caption{\label{table4}
Calculated final-state contributions to the total SCLSs of first layer Ru atoms directly coordinated to zero, one, two or three adsorbed H atoms.}
\begin{tabular}{@{}l|cccc}
\hline
                                 & 0fold & 1fold & 2fold & 3fold \\ \hline
Clean                            & $-$62 &       &       &       \\
$(2 \times 2)$-O                 & $-$45 &       &       &       \\
$(2 \times 2)$-(O+H$_{\rm fcc}$) & $-$75 & $-$36 &       &       \\
$(2 \times 2)$-(O+3H$_{\rm hcp}$)&       &       & $-$38 & $-$38 \\
$(2 \times 2)$-4H$_{\rm fcc}$    &       &       &       & $-$44 \\
\end{tabular}
\end{table}

As a first step to analyze the physics governing the observed additivity rule it is useful to decompose the total calculated shifts into initial- and final-state contributions. Table~\ref{table4} compiles the final-state contributions to all first layer SCLSs of the five measured structures. Overall, the screening correction is found to be of the same order of magnitude as the actual total H-induced shifts presented in Fig.~\ref{fig4}, as well as only marginally smaller than the screening contributions reported in ref. \cite{lizzit01} for the O-induced SCLSs at Ru(0001). This is distinctly different to the situation for the initial-state shifts, which for the latter O-containing structures were one order of magnitude larger than the now considered H-induced shifts.

The consistently obtained negative sign of the screening correction indicates that the screening capabilities of the core-ionized system are enhanced at the surface compared to the bulk. For the clean surface, the detailed analysis put forward in refs. \cite{lizzit01,andersen94} has already rationalized this by the narrowing of the local valence $d$-band induced by the lowered coordination of the surface atoms. This in turn leads in general to an enhancement of the density of states (DOS) at and above the Fermi level compared to the bulk situation, which $-$ with the final-state effects at a transition metal surface like Ru(0001) largely due to the intra-atomic $d$-electron screening \cite{andersen94} $-$ implies a more efficient screening at the surface. That this is not much affected by the presence of the adsorbates can be understood considering the Anderson-Grimley-Newns model of chemisorption \cite{grimley75}: Consistent with this model the computed adsorbate induced DOS-change concerns primarily the formation of a bound state at the bottom of the substrate $d$-band and a broad (antibonding) resonance at the upper edge of the $d$-band. With the Fermi-level in Ru positioned at about 2/3 band-filling, this leaves the DOS at the Fermi-level and thereby the screening capabilities largely unaffected compared to the situation at the clean surface. In this understanding one would also not expect a strong influence of the adsorbate induced geometric relaxation on the magnitude of the SCLS screening contribution. Corresponding calculations in which the Ru atoms were frozen to their positions in a bulk-truncated surface geometry and only the adsorbate positions were fully relaxed indeed confirm this expectation with computed final-state shifts that differ by less than 10\,meV from the corresponding values for the fully relaxed geometries summarized in Table~\ref{table4}. Overall, the final-state contributions obtained for the O-induced shifts and H-induced shifts are thus quite similar, with the values for different H coordinations showing even less variations. With a magnitude of $\sim -55 \pm 20$\,meV, consideration of these final-state effects is essential to reach the reported quantitative agreement with the measured total SCLSs. On the other hand, the scatter obtained in the final-state contributions for the different H coordination situations is too small to obscur the trend of about linearly increasing initial-state shifts with increasing H coordination, even though the latter are much smaller than the O-induced ones where the additivity rule was observed before.

As expected from the understanding developed in ref. \cite{lizzit01} we therefore arrive at the conclusion that the observed correlation of first layer SCLSs with the number of H nearest neighbors is an initial-state effect. The formation of bonds between the adsorbate and the first layer Ru atoms affects the local valence $d$-band, the center of which shifts to maintain local charge neutrality. This changes the local (near nucleus) electrostatic field, which in turn acts on the core electrons as well and induces the finite SCLSs. Effectively, the about linear dependence of the latter on the number of H neighbors thus simply reveals that the type of bonding, i.e. the bonds formed per H nearest neighbor, is roughly the same in all overlayers studied. With all caveats given in ref. \cite{lizzit01} this may be interpreted in the conceptual language of charge transfer in that the amount of charge transferred to each H atom remains approximately constant in the different structures. The much smaller electronegativity of H then also rationalizes the much smaller observed H-induced initial-state shifts ($\sim 50$\,meV per H atom) compared to the O-induced shifts reported before ($\sim 400$\,meV per O atom \cite{lizzit01}), and is fully consistent with the very different work function changes induced by the two adsorbates. In agreement to experiment \cite{feulner85} we compute the latter to be essentially zero even for the full monolayer H coverage, in stark contrast to measured \cite{madey75} and computed \cite{reuter02} O-induced changes of up to $\sim 1.5$\,eV for the full $(1 \times 1)$-O monolayer. The thereby suggested near neutrality of H on Ru(0001) may also explain why the additivity rule prevails even in the denser adsorbate overlayers. For an effectively charged adsorbate, some degree of depolarization, i.e. reduced charge transfer per adsorbate, would be favorable for higher coverages as it reduces the repulsive adsorbate-adsorbate interaction. This effect is indeed observed for dense O overlayers \cite{pirovano01,lizzit01}, but is too small there to significantly affect the near-linear SCLS correlation with O coordination. In this respect, one would expect a noticeable violation of the additivity rule particularly for adsorbates with only a modest bond strength (implying smaller total magnitudes of the adsorbate-induced initial-state SCLSs), but relatively strong induced dipole moment (suggesting more pronounced depolarization effects at decreased adsorbate-adsorbate distances). This situation could for example be given for molecular adsorbates like CO or NO, and we are currently analyzing coadsorbate overlayers containing O, H and these molecular species in this regard \cite{sclstobepublished}. 

With the SCLSs sensitively reflecting the formed adsorbate-substrate bonds, it is clear that they can also be used as local probes of the adsorption site. This was already shown in the preceding work on the O-containing structures, where the computed Ru SCLSs for structures with O atoms in fcc sites were consistently about 100-200 meV smaller than the corresponding values in structures with O atoms in hcp sites and therewith irreconcilable with the experimental data \cite{lizzit01}. For O adsorption into hcp sites the induced SCLS change per O neighbor is thus slightly larger, whereas, interestingly, for Rh(111) where the fcc site is the more stable adsorption site the situation is exactly reversed in that there O adsorption into fcc sites leads to slightly larger SCLS increases \cite{pirovano01}. For both surfaces, a somewhat more ``negatively charged'' oxygen atom is thus adsorbed in the more stable adsorption site, suggesting a preference for a stronger ionic bonding. In this respect, it is intriguing to notice from Fig.~\ref{fig4} that also for the H overlayers the slightly more stable fcc site is the one that leads to slightly larger induced SCLS changes. The less favorable hcp site is only populated in the O-containing dense $(2 \times 2)$-(O+3H) coadsorbate structure and it is tempting to attribute this stabilization in parts to that a slightly less ionic hydrogen is preferred at the corresponding short inter-adsorbate distances. While in this structure the effect concerns only the position of the H atoms with the O atoms remaining in the hcp sites, coadsorption of CO is able to shift also the O atoms to fcc sites. Analyzing the correlation of the respective stabilities of different sites with the induced SCLS changes is thus another aspect of the mentioned on-going work on coadsorption structures containing O, H, CO and NO \cite{sclstobepublished}.

\section{Conclusions}

In conclusion we have presented a joint experimental and theoretical study of Ru(0001) surface core level shifts induced by O and H adsorbates. Remarkable quantitative agreement concerning the energetics, geometries, and SCLS values of the investigated 
$(1 \times 1)$-H, $(2 \times 2)$-(O+H), and $(2 \times 2)$-(O+3H) (co)adsorbate overlayers on Ru(0001) is achieved. The quasi-linear dependence of the adsorbate-induced SCLSs on the number of directly coordinated nearest neighbors reported before for O overlayers on Ru(0001) \cite{lizzit01}, Rh(111) \cite{pirovano01} and Rh(100) \cite{baraldi04}, as well as for H adsorbate layers on all Rh low-index single crystal surfaces \cite{baraldi08} is found to prevail also in the H-(co)adsorbate structures on Ru(0001). The computed decomposition of the total SCLSs into initial- and final-state contributions reveals an enhanced screening at the surface of the (co)adsorbate layers that is of the same order of magnitude as for the pure O overlayers studied before. Accounting for this final-state correction is thus necessary to obtain the reported quantitative agreement with the measured data, while its actual value exhibits only a small scatter for the differently coordinated surface atoms. In accordance with the detailed analysis put forward in ref. \cite{lizzit01}, the observed additivity of the adsorbate-induced SCLSs is thus a pure initial-state effect. In terms of the conceptual charge transfer language it merely reflects a roughly contant amount of charge transferred to each directly coordinated adsorbate neighbor. Not surprisingly, this amount of charge (and concomitantly the magnitude of the observed SCLS change) is significantly smaller for the almost electroneutral H adsorbates than for the electronegative O adsorbates. Intriguingly, the SCLS data reveals slightly larger induced changes for H atoms adsorbed in fcc sites than in hcp sites, consistent with the interpretation that a slightly more ionic H adsorbs in the prior sites. The established sensitivity even to such highly similar adsorption sites, as well as the prevalence of the additivity rule confirms the suggested use of SCLSs as directly accessible and valuable fingerprints of the adsorbate populations.

\section*{Appendix}

Similar to the analysis performed in ref. \cite{lizzit01}, the experimental data were fitted with a combination of Doniach-$\breve{\rm{S}}$\`unji\'c functions \cite{doniach70} convoluted with a Gaussian broadening and a linear background. The lineshape parameters are the Lorentzian ($L$), Gaussian ($G$) and asymmetry ($\alpha$). The following outlines the fitting procedure used to extract these parameters and the values of the Ru $3d_{5/2}$ SCLSs.

The absolute binding energy of the peaks was calibrated to the Fermi-level, measured for each spectrum. Moreover, the intensity of the spectra was normalized to the background at the low binding energy side (279.5\,eV) in order to account for variations of the photon beam intensity. The best fit parameters were extracted from the fit of all the data taken at the three photon energies used. First the spectrum of the clean surface was fitted with three components: bulk ($L = 0.20$\,eV, $G = 0.06$\,eV, $\alpha = 0.07$), S$_1$ ($L =0.24$\,eV, $G = 0.06$\,eV, $\alpha = 0.05$), and S$_2$ ($L = 0.20$\,eV, $G = 0.06$\,eV, $\alpha = 0.08$). The absolute position of the bulk peak was determined as well and this position was kept fixed in all other fits. Next, we fitted the $(2 \times 2)$-O spectrum to determine the S$_1$(O) lineshape ($L = 0.24$\,eV, $G = 0.12$\,eV, $\alpha = 0.06$), which is wider than the S$_1$ component of the clean surface as already found in our previous work \cite{lizzit01}. The S$_1$ peak intensity turns out to be $\sim 1/4$ of the same peak of the clean surface, as it should be at the oxygen coverage of 0.25\,monolayer. Moreover, the intensity of the bulk peak is increased while that of S$_2$ is lowered, consistent with the expectation that a near-zero and unresolved S$_2$(O) peak contributes to the bulk region.

We then continued with the $(1 \times 1)$-H structure where again three components are expected: bulk, S$_2$ (same lineshape parameters as S$_2$ in the clean surface) and S$_1$(3H$_{\rm fcc}$) ($L =0.24$\,eV, $G = 0.10$\,eV, $\alpha = 0.05$). It has to be noted here that the intensity of the different components is almost the same as for the clean surface, because the hydrogen atoms present on the surface have only small scattering effects on the electrons photoemitted from the Ru substrate. On the other hand, the presence of H at the surface seems to affect the peak shape in the spectra, and we obtain significantly improved fits when allowing for a larger Gaussian width. As mentioned in the text, we interpret this by inhomogeneous broadening due to a less than perfect ordering achieved for the H overlayer. The same situation (unchanged intensity with respect to the $(2 \times 2)$-O, increased Gaussian width) is found for the coadsorbate structures, and the line shape parameters determined are: S$_1$(O+1H$_{\rm fcc}$) ($L = 0.24$\,eV, $G = 0.15$\,eV, $\alpha = 0.06$) in the $(2 \times 2)$-(O+H) structure; and S$_1$(O+2H$_{\rm hcp}$) ($L = 0.26$\,eV, $G = 0.16$\,eV, $\alpha = 0.06$) and  S$_1$(3H$_{\rm hcp}$) ($L = 0.24$\,eV, $G = 0.12$\,eV, $\alpha =0.05$) in the $(2 \times 2)$-(O+3H) structure.

\end{document}